\documentclass{cimento}

\usepackage{graphicx}

\title{Theoretical interpretation of GRB 011121}
\author{C.L.~Bianco\from{icra}\from{dipfis},
L.~Caito\from{icra}\from{dipfis},
\atque
R.~Ruffini\from{icra}\from{dipfis}}

\instlist{
  \inst{icra} ICRANet and ICRA, Piazzale della Repubblica 10, I--65100 Pescara, Italy.
  \inst{dipfis} Dipartimento di Fisica, Universit\`a di Roma ``La Sapienza'', Piazzale Aldo Moro 5, I-00185 Roma, Italy.
  }
\PACSes{\PACSit{98.70.Rz}{gamma-ray sources; gamma-ray bursts}}

\begin{document}

\maketitle

\begin{abstract}
GRB011121 is analyzed as a prototype to understand the ``flares'' recently observed by Swift in the afterglow of many GRB sources. Detailed theoretical computation of the GRB011121 light curves in selected energy bands are presented and compared and contrasted with observational BeppoSAX data.
\end{abstract}

\section{Introduction}

On the 21 November 2001 at 18:47:21 UT, GRB 011121 was observed by BeppoSAX satellite \cite{1}. The localization was then refined by the Interplanetary Network \cite{2} and an optical transient was identified by the OGLE group \cite{3}. Spectroscopic measurements revealed some emission lines at the redshift z = 0.36 \cite{4}.

GRB 011121 is a long burst with $T_{90} = 28$ s; the fluence \cite{5} is $2.4\times 10^{-5}$ erg/cm$^2$ that corresponds, in the hypothesis of isotropic emission and at the considered redshift, to an energy in the band $2$-–$700$ keV of $2.8\times 10^{52}$ erg. This is the second brightest source detected by BeppoSAX both in $\gamma$-rays and in X-rays \cite{6}.

By the observation of the light curves and spectra in different energy bands, we can note \cite{6,7} a flare, at the time $t_a^d = 240$ s. It lasts approximately seventy seconds and corresponds to a bump of an order of magnitude in luminosity. It is however very soft, since its energy is about the 3\% of the total amount of the prompt emission \cite{6}. We here show how we can reproduce the flare as due to a spherical cloud in the interstellar medium (ISM) along the line of sight of the GRB source \cite{8,11}.

\section{The fit of the GRB 011121 observed luminosity}

\begin{figure}
\includegraphics[width=0.5\hsize,clip]{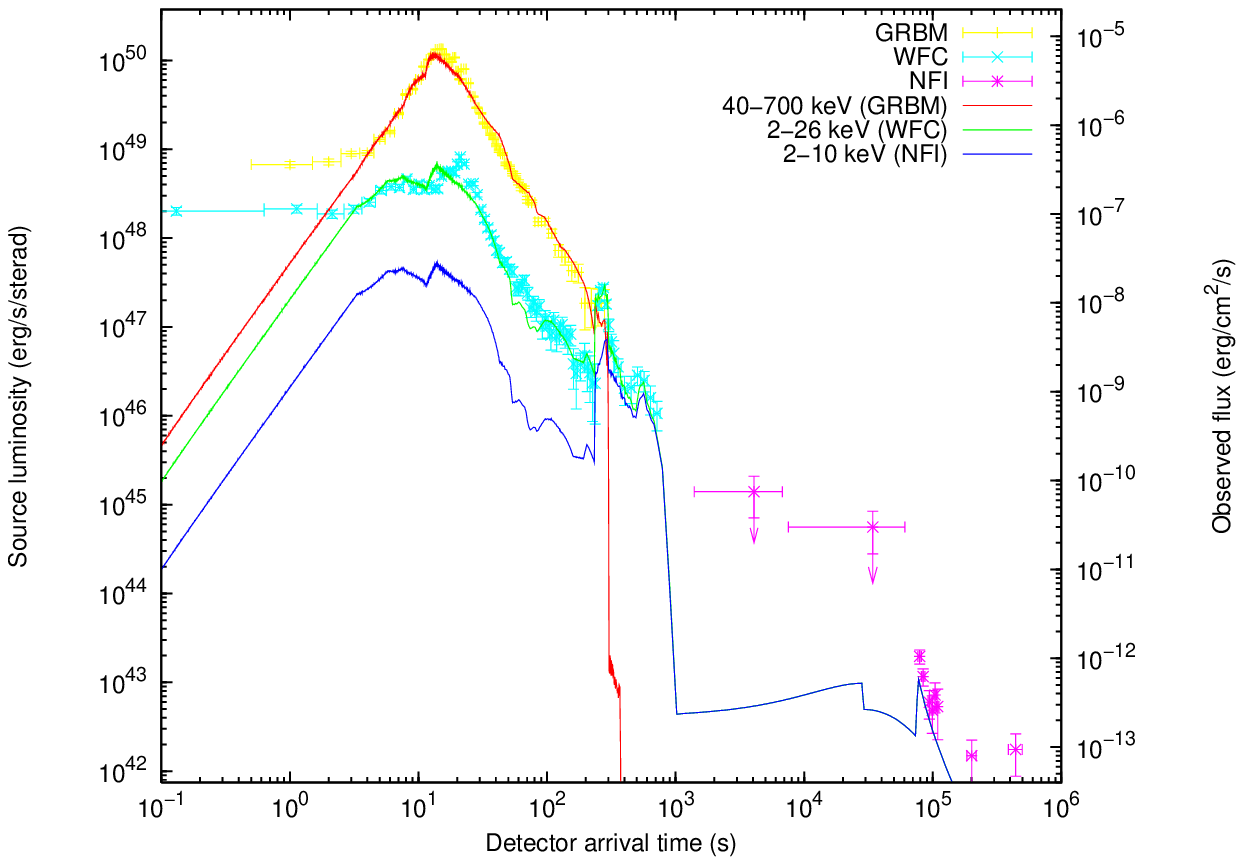}
\includegraphics[width=0.5\hsize,clip]{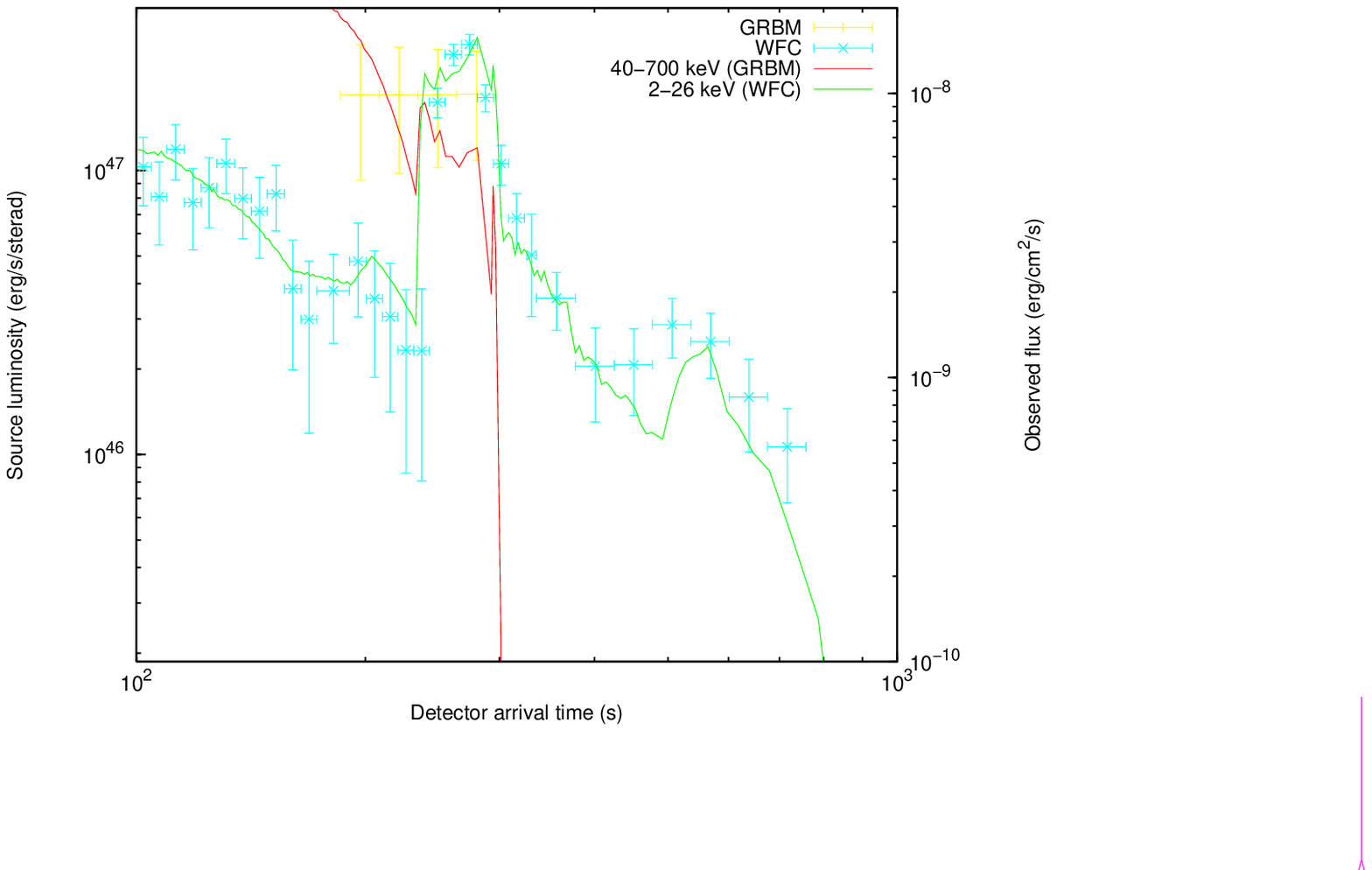}
\caption{{\bf Left panel:} Theoretical fit of the GRB 011121 light curves in the $40$-–$700$ keV (BeppoSAX GRBM), $2$-–$26$ keV (BeppoSAX WFC), $2$-–$10$ keV (BeppoSAX NFI). {\bf Right panel:} Enlargement of the flare.}
\label{fig}
\end{figure}

In left panel of Fig. \ref{fig} we present the observed GRB 011121 light curves in the three different energy bands we analyzed, together with their theoretical fit in the framework of our model (see Refs. \cite{8,9,10,11,12,13} and references therein): $40$-–$700$ keV, $2$-–$26$ keV, $2$-–$10$ keV.

Looking at the observational data we can see that the $40$-–$700$ keV light curve presents a temporal profile particularly regular, smooth and homogeneous, while the $2$-–$26$ kev light curve has a remarkably irregular profile.

In the right panel of Fig. \ref{fig} there is an enlargement of the flare, showing in details the comparison between the theoretical light curve and the observational data. We recall that in our model the afterglow emission is due to the interaction with the ISM of the shell of baryons accelerated during the optically thick plasma expansion phase \cite{8,9}. We can see that the agreement is very good, even in the decaying tail of the flare. In the computation of the theoretical light curve we have, for the first time, taken into account the three-dimensional structure of the ISM, instead of assuming a purely radial density profile \cite{8,11}.

\end{document}